\documentclass[aps,prl,amsmath,amssymb,twocolumn,showpacs,superscriptaddress]{revtex4}
\usepackage{graphicx}

\usepackage{epstopdf}
\begin{document}

\author{Emanuela Zaccarelli}
 \affiliation{Dipartimento di Fisica and INFM-CRS-SOFT, Universit\`a
  di Roma {\em La Sapienza}, Piazzale A. Moro 2, 00185 Roma, Italy }
\author{Stefan Andreev}
\affiliation{Department of Chemistry, Columbia University, 3000
  Broadway, New York, New York 10027, USA}
\author{Francesco Sciortino}
\affiliation{Dipartimento di Fisica and INFM-CRS-SOFT, Universit\`a
  di Roma {\em La Sapienza}, Piazzale A. Moro 2, 00185 Roma, Italy }
\author{David R. Reichman}
\affiliation{Department of Chemistry, Columbia University, 3000
  Broadway, New York, New York 10027, USA}

\title{Numerical Investigation of Glassy Dynamics in Low Density Systems}

\date{\today}

\begin{abstract}
Vitrification in colloidal systems typically occurs at high densities
driven by sharply varying, short-ranged interactions. 
The possibility of glassy behavior arising from
smoothly varying, long-ranged particle interactions has received
relatively little attention.
Here we investigate the behavior of 
screened charged particles,
and explicitly demonstrate that these systems exhibit glassy
properties in the regime of low temperature and low density.
Properties close to this low density (Wigner) glass transition share
many features with their hard-sphere counterparts, but differ in
quantitative aspects that may be accounted for via microscopic
theoretical considerations.
\end{abstract}

\pacs{64.70.Pf, 61.20.Lc, 82.70.Dd}

\maketitle

The origins of the precipitous slowing down of dynamics in supercooled
liquids are still unclear even after many decades of intense scrutiny.
Model systems often form the basis for detailed investigations that
include the core features known to give rise to generic glassy
behavior.  Perhaps the most prominent example of such a model system
is the hard-sphere suspension, which has served as the basis for
numerous experimental, theoretical and computational studies of the
glass transition\cite{Pus86a,Bar89a,Voi04a}.  Many of the most
interesting properties of supercooled liquids and glasses, including
two-step relaxation, stretched exponential decay of density
correlations and dynamic heterogeneities occur in
hard-sphere systems\cite{VanM93,Wee02,Bert05}.

While the glass transition of hard-spheres has become a
paradigm for vitrification at {\em high densities}, serving as a
reference point for conceptual attempts to connect the behavior of
physically diverse classes of disordered arrested states of matter,
another physically important limit of glassy systems has received much
less systematic scrutiny.  This limit is that of a dilute assembly of
particles interacting via long-ranged, soft repulsive forces such as
those arising in charged systems.  Over twenty years ago Chaikin and
coworkers\cite{Cha1982,Cha1989} investigated the phase behavior of
dilute suspensions of charged colloids. Low density glasses stabilized
by Coulomb repulsion were called "Wigner glasses"\cite{Cha1982}. At
that time a detailed investigation of the structure and dynamics of
these suspensions was not carried out.  Hints of glassy behavior in
one component plasmas have also been noted\cite{Ichi82}, and the
notion of a Wigner glass has been revived in colloidal
systems\cite{Bon99b,Bec99a,Sci04a,ChenPRE} due to recent activity focusing on physical gelation in charged
systems\cite{Ruz04,Schurt,Cha07a,Mos07a,Cab07a}. Indeed, the study of
glassy properties of dilute Coulomb systems has consequences that
reach beyond classical systems and may shed light on routes to the
formation of glasses in electronic systems\cite{Ova97,Ova04}, where
glassy effects might persist even in the limit of weak to vanishing
quenched disorder\cite{Mull04}. Such self-induced glassiness results
from electron-electron interactions, an effect analogous to classical
Wigner glass formation in colloidal systems.

Similar to the experimental situation, few theoretical investigations
of the emergence of glassy dynamics in low density Coulomb systems
have been carried out\cite{Ros89a}. The most detailed investigations
have been performed by Bosse and Wilke\cite{Bos98a,Wil99a}, who have
used idealized mode-coupling theory (MCT) \cite{goetze,barrat} to
predict the dynamical behavior of a low density charged system in a
neutralizing background as it approaches the putative Wigner glass
transition.  These authors predict that glassy behavior in this dilute
regime shares many properties with the high density hard-sphere
system, but some unique behavior also emerges. Indeed, Wigner glasses
are stable even in the extreme dilute limit where the static structure
factor shows no modulation due to molecular shell structure.
Furthermore, when the electrostatic repulsion is complemented by an
excluded volume interaction, the MCT glass line is predicted to show a
reentrant behavior due to the (density dependent) competition between
the hard-core and the soft long-ranged repulsion\cite{Lai97a,Bos98a}.

There are two main goals for the work presented here.  First, we show
that it is possible to generate Wigner glasses in simulations of
dilute binary long ranged (screened) Coulomb system with a judicious
choice of the interaction potential parameters and of the studied
state point.  Indeed, no previous theoretical work has addressed the
stability of the Wigner glass with respect to the competing facility
of crystallization\cite{Rob88a,Mei91a,Low91a,Horb07}. This first
result has important implications for the modeling and interpretation
of dynamics in charged colloidal suspensions\cite{Roy03a,RoyJCP}.  Our
second goal is to systematically study the glassy behavior of dilute
Coulomb systems and to test some aspects of the MCT predictions for
dynamics as the Wigner glass state is approached.

Our choice of interparticle potential is dictated by a reasonable
compromise between simplicity and realism. In particular, our
simulations employ the standard Yukawa form.
Due to the rapid crystallization of one-component Yukawa systems, we
have investigated $50\%-50\%$ binary mixtures of 
point particles that may serve as
models for dilute Wigner glasses.  It should be noted that we have not
included an additional short-range hard-sphere term in the potential.
The advantage of this (unrealistic) simplification is that it allows
us to study the effects of soft, long-ranged interactions in isolation
of hard-core contributions.  A disadvantage of this choice is that the
notion of volume fraction is ill-defined, hence we will discuss our
results in terms of number density $n=N/L^3$ where $L$ is the
simulation box length and $N$ is the number of particles. The issue of reentrance, crucially
depending on the existence of disparate length scales, may only be
investigated if the hard-core repulsion is included
and will be the subject of future
work.

The interaction potential between
species $i$ and $j$ is,
\begin{equation}
V_{ij}(r)=A_{ij}\frac{\exp(-r/\xi_{ij})}{r/\xi_{ij}}.
\end{equation}
The amplitude of the repulsion differs for the two species as $A_{11}=0.2\epsilon$, $A_{22}=3.5\epsilon$, and $A_{12}=0.837\epsilon$, with $\epsilon$ being the unit of energy, 
while the screening lengths are all identical and taken to be
$\xi_{ij}=\xi=1$.  Lengths are measured in units of $\xi$.
Such mixture corresponds to a model of two dilute colloidal species
with different surface charges\cite{notamixt}.
We perform molecular dynamics simulations of point particles of unitary mass $m$ for two system syzes, i.e. $N=10^3$ and $N=10^4$, at a fixed number density $n=0.002984$, while the
temperature $T$ (measured in units of $k_B$, which we set equal to 1)
is varied. In the following we always use the reduced temperature
$T^*=T/10^{-5}$. The time step is 0.5 in units of $\xi/(m \epsilon)^{1/2}$. 
Since the potential is long-ranged, we choose a cutoff distance equal
to $25\xi$, which ensures that all
repulsive contributions at sufficiently long distances are accounted for. For all studied state points with $T^* \geq 1.7$ data has been collected in the NVE ensemble following a sufficiently long equilibration period, controlling that no aging phenomena are detectable.  Simulations for $T^*\lesssim 1.7$ show 
aging even after extremely long equilibrations, confirming the approach to a non-ergodic state.

\begin{figure}
\includegraphics[width=6.5cm, clip=true]{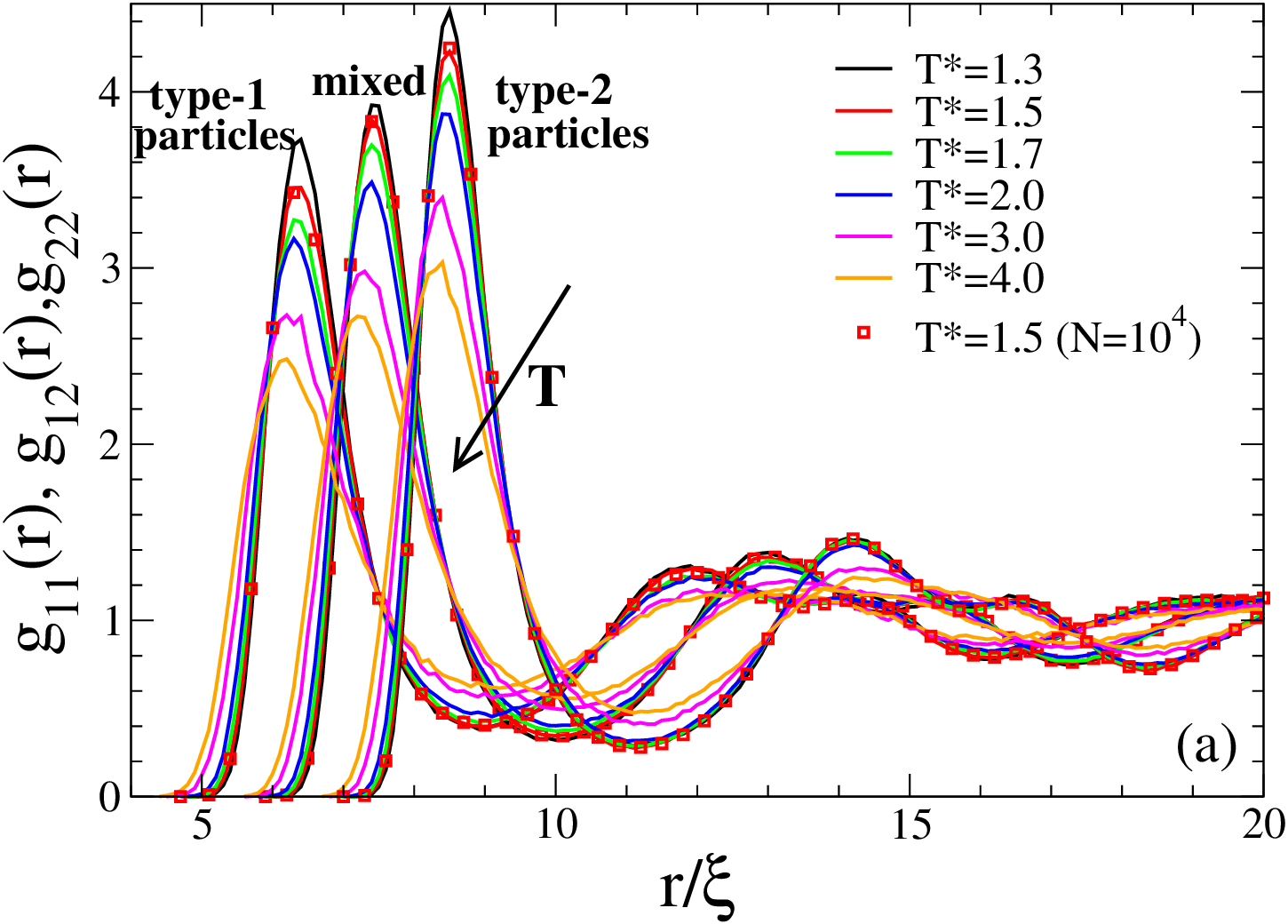}
\includegraphics[width=6.3cm, clip=true]{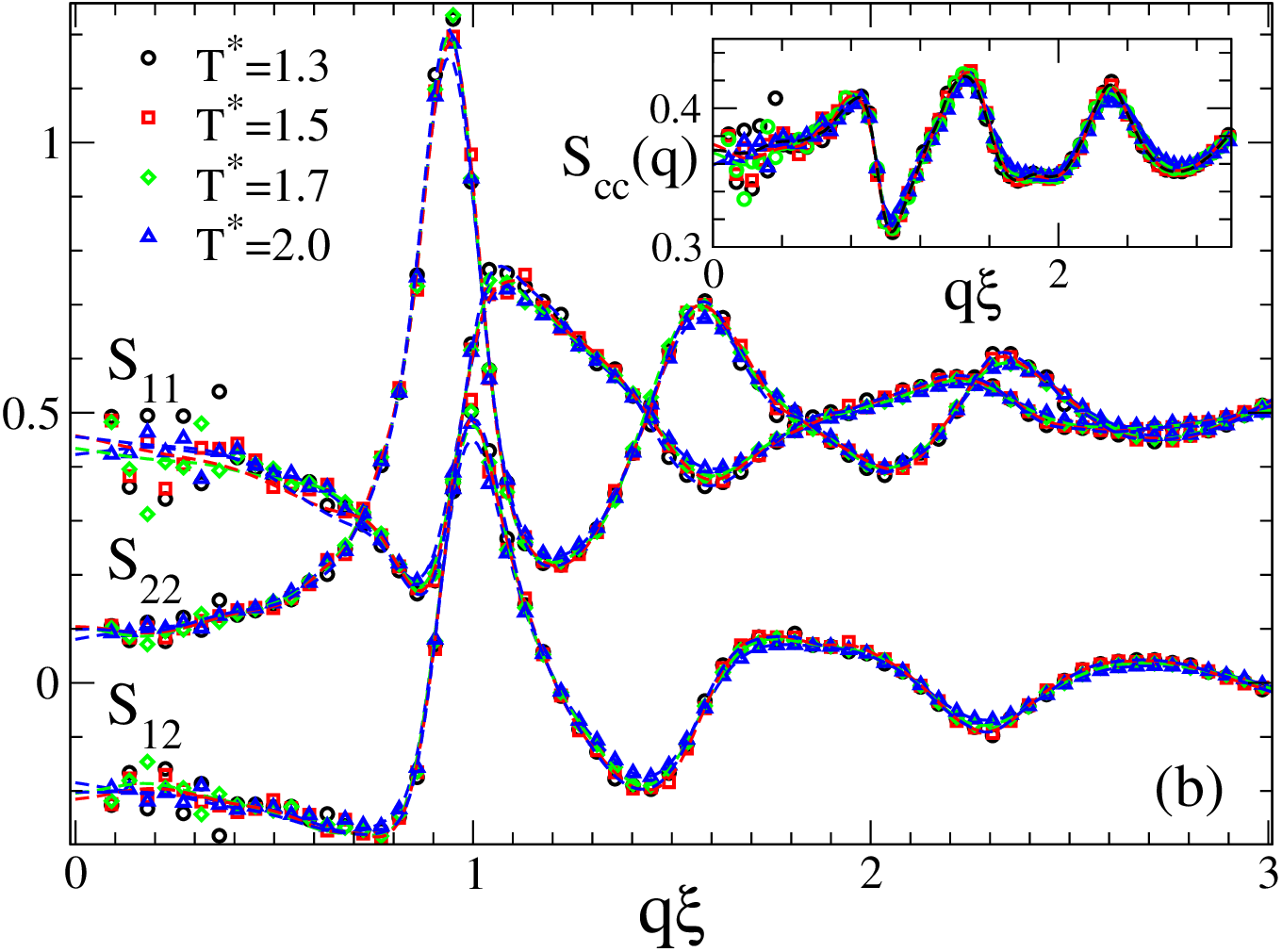}
\caption{
Evolution in the supercooled regime, upon decreasing $T$, of: (a)
partial radial distribution functions $g_{11}(r)$, $g_{12}(r)$ and $g_{22}(r)$. The
perfect agreement between two different simulation sizes,
i.e. $N=10^{3}$ (lines) and $N=10^{4}$ (symbols), provides evidence
that results are equilibrated, reproducible and not phase-separating; 
(b) partial static structure factors $S_{ij}(q)$, as
well as $S_{cc}(q)$\protect\cite{Bha1970} which rules out the presence
of a demixing transition. Symbols are results for simulations with
$N=10^{4}$ particles, lines are splined curves. }
\label{fig:sq}
\end{figure}

In Fig.\ref{fig:sq} we show results for the partial radial
distribution functions $g_{ij}(r)$  and for the partial static
structure factors $S_{ij}(q)$ in the supercooled regime.  A
progressive structuring, as well as a shift towards larger distances,
of the peaks of all partial radial distribution functions is observed upon decreasing
$T$. Standard signatures of supercooled liquid
structure, such as a split in the second peak, are evident.  Type-$1$
particles are significantly closer on average than type-$2$ ones, due
to the weaker repulsion.  No sign of crystallization is noted during
the whole duration of the runs.  Focusing on the partial structure
factors, we notice that the amplitude of all $S_{ij}$ is small as compared to
the canonical glass formers. Moreover, $S_{11}(q)$ shows a consistent upturn at low $q$, again a manifestation of the smaller repulsion experienced by the
type-$1$ particles with respect to the type-$2$ particles.  Indeed,
such feature is less evident for $S_{12}$ while it is absent  for $S_{22}$.  To rule out the
possibility that a phase separation process interferes with the
dynamical slowing down, we have monitored
the time dependence of $S_{11}(q)$ and visually inspected the
configurations, both observations providing evidence against this
putative process. Moreover, upon further decrease of $T$, the low-$q$
amplitude of $S_{11}(q)$ saturates to a constant value.  The
concentration-concentration structure factor
$S_{cc}(q)$\cite{Bha1970}, (shown in the inset of Fig.\ref{fig:sq}(b))
does not display any increase at low $q$, ruling out the
possibility of a demixing transition.   Results for the
structural properties $g_{ij}(r)$ and $S_{ij}(q)$ for the larger
system size, also shown in Fig. \ref{fig:sq}, are identical to those
of the smaller system.

\begin{figure}
\includegraphics[width=6.5cm, clip=true]{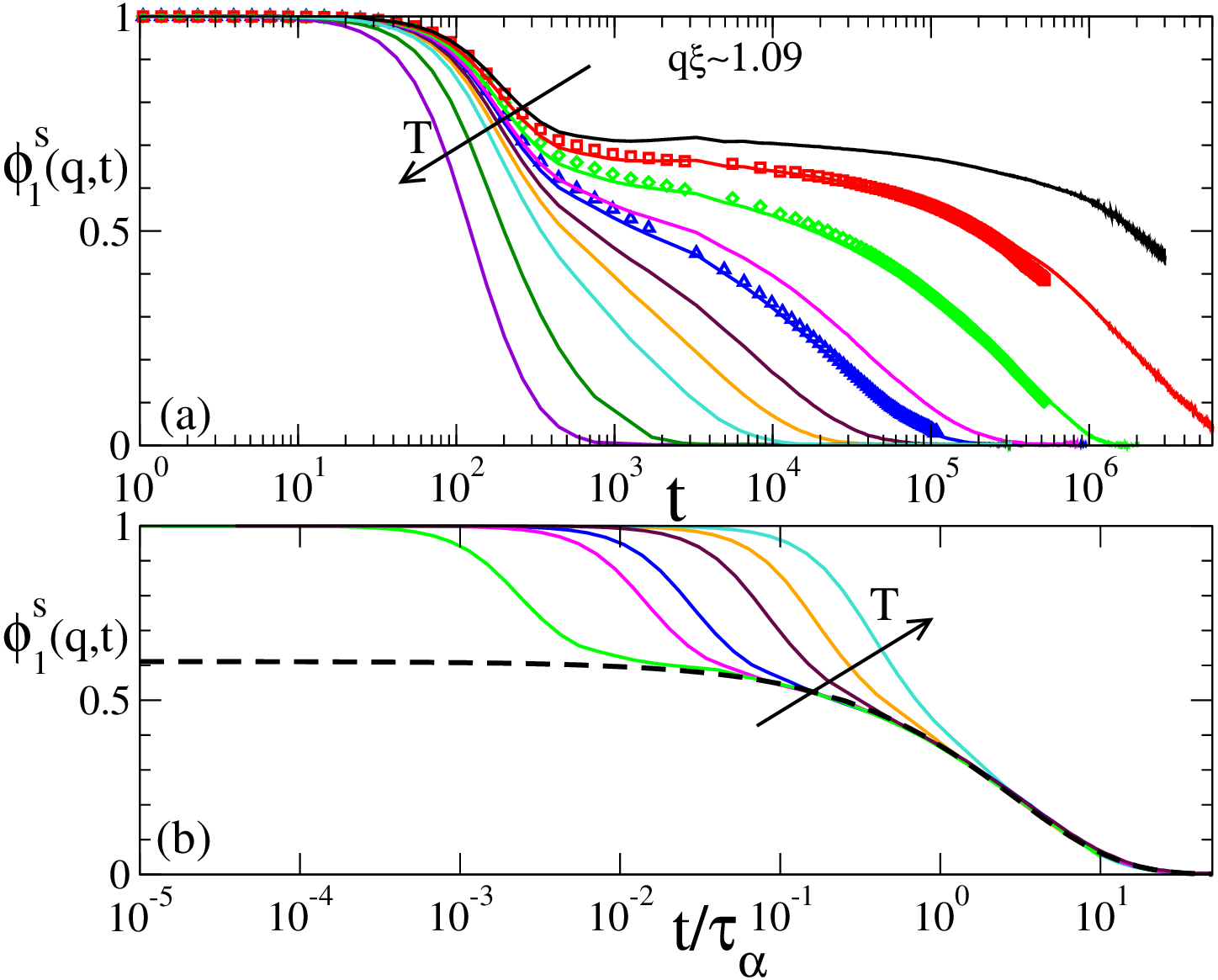}
\caption{
Self autocorrelation function for the density of type-$1$ particles
$\phi^s_1(q,t)$ at $q\xi\simeq1.09$: (a) $T$ dependence (from left to right: $T^*=10, 5, 3, 2.5, 2.25, 2, 1.9,1.7, 1.5, 1.3$) and (b) TTS
scaling only in the temperature range $1.7 \lesssim T^* \lesssim 3$. The dashed line is a stretched exponential fit of the master
curve with stretching exponent $\beta\sim 0.65$.  In (a) data for the
larger simulation size (symbols) are also reported.  Below $T^*_c=1.7$ the
system shows a clear aging behavior, as well as a breakdown of TTS due
to the increase in the plateau height. For type-$2$ particles, a
similar, albeit slower dynamics (due to the larger repulsion amplitude), occurs.}
\label{fig:self}
\end{figure}
Having demonstrated that our system shows stable structural behavior,
we next turn to investigate the dynamical behavior.  In
Fig.~\ref{fig:self}-a we show the decay of the self autocorrelation
function of the density of type-$1$ particles, $\phi^s_{1}(q,t)$
close to the first peak of the 
structure factor.
Strikingly, the behavior observed is qualitatively similar to that
seen in the familiar {\em dense} systems containing a harsh repulsive
core. In particular, the dynamics slows down by several orders of
magnitude, giving rise to a typical two-step decay, in which the final
relaxation is of stretched exponential form.  Hence, the formation of
a wide, soft cage around each particle arises, due to the {\em long-range} repulsive
interactions of different strengths.

In Fig.~\ref{fig:self}-b 
we test  time-temperature superposition (TTS) via rescaling
$\phi^s_1(q,t)$ by the ($T$-dependent) $\alpha$-relaxation time.  For
all $\phi^s_1(q,t)$ showing a two-step decay TTS holds down to
$T^*=1.7$.  For $T^*\lesssim 1.7$ the plateau height changes
significantly, leading to a clear breakdown of TTS. In the same
$T$-range, aging effects can be observed, namely extremely slow drifts
of the energy as well as dependence of the dynamical variables of the
initial observation times.

\begin{figure}[tbh]
\includegraphics[width=6.5cm, clip=true]{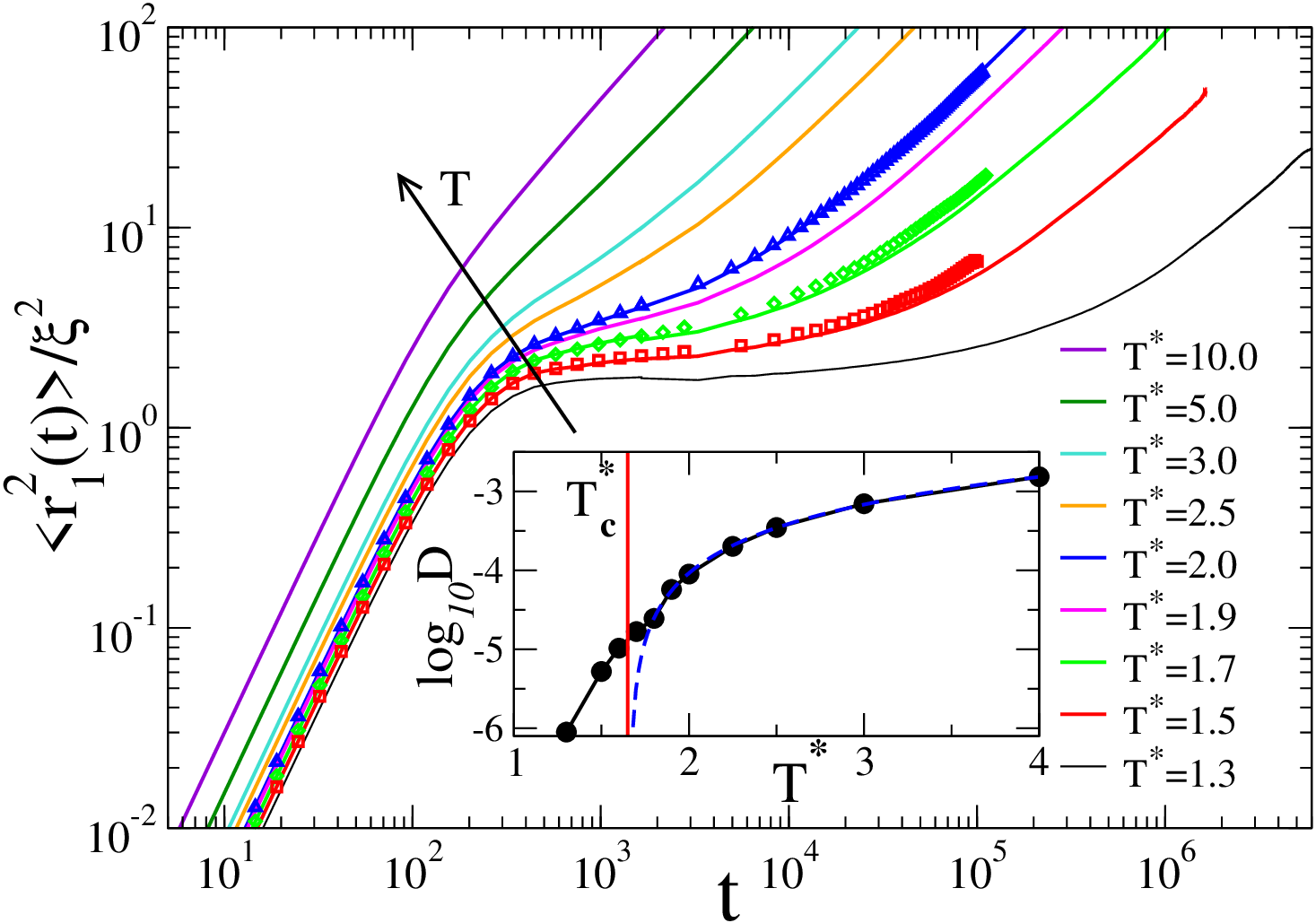}
\caption{
$\langle r^2_1(t)\rangle$ for type-$1$ particles with decreasing $T^*$. Lines and
symbols refer respectively to simulations with $N=10^3$ and $N=10^4$
particles.  Inset: Power-law fit to the diffusion coefficient $D$,
with exponent $\gamma_D=1.45$ and $T^*_c\simeq1.67$.}
\label{fig:msd}
\end{figure}

We also study the mean squared displacement $\langle r^2_1(t)\rangle$
as a function of $T^*$.  It shows the typical development of a long
intermediate plateau, indicative of a dynamical slowing down (see
Fig. \ref{fig:msd}). The magnitude of the plateau is quite high,
larger than $\xi$.  
The extracted diffusion coefficient $D$, evaluated from the long time limit
of the MSD, is reported in the inset.  Power-law fits to $D \sim | T^*-T_c^*|^{\gamma_D}$ provide a method to estimate an effective "mode-coupling
temperature" of $T_c^*= 1.7$, although the extracted exponent $\gamma_D=1.45$  is smaller than the lowest limit set by MCT\cite{goetze}.
Below $T^*_c$,  we observe 
clear deviations from the power-law behavior
suggesting  that hopping processes are particularly
relevant due the softness of the repulsive cages.  A further evidence of 
a change in the dynamics around $T_c^*$ is the fact that 
 the  plateau height,  which can be operatively defined as the inflection point of
$\log(MSD)$ vs. $\log(t)$,  significantly decreases for $T^*< 1.7$.

\begin{figure}[tbh]
\includegraphics[width=6.5cm, clip=true]{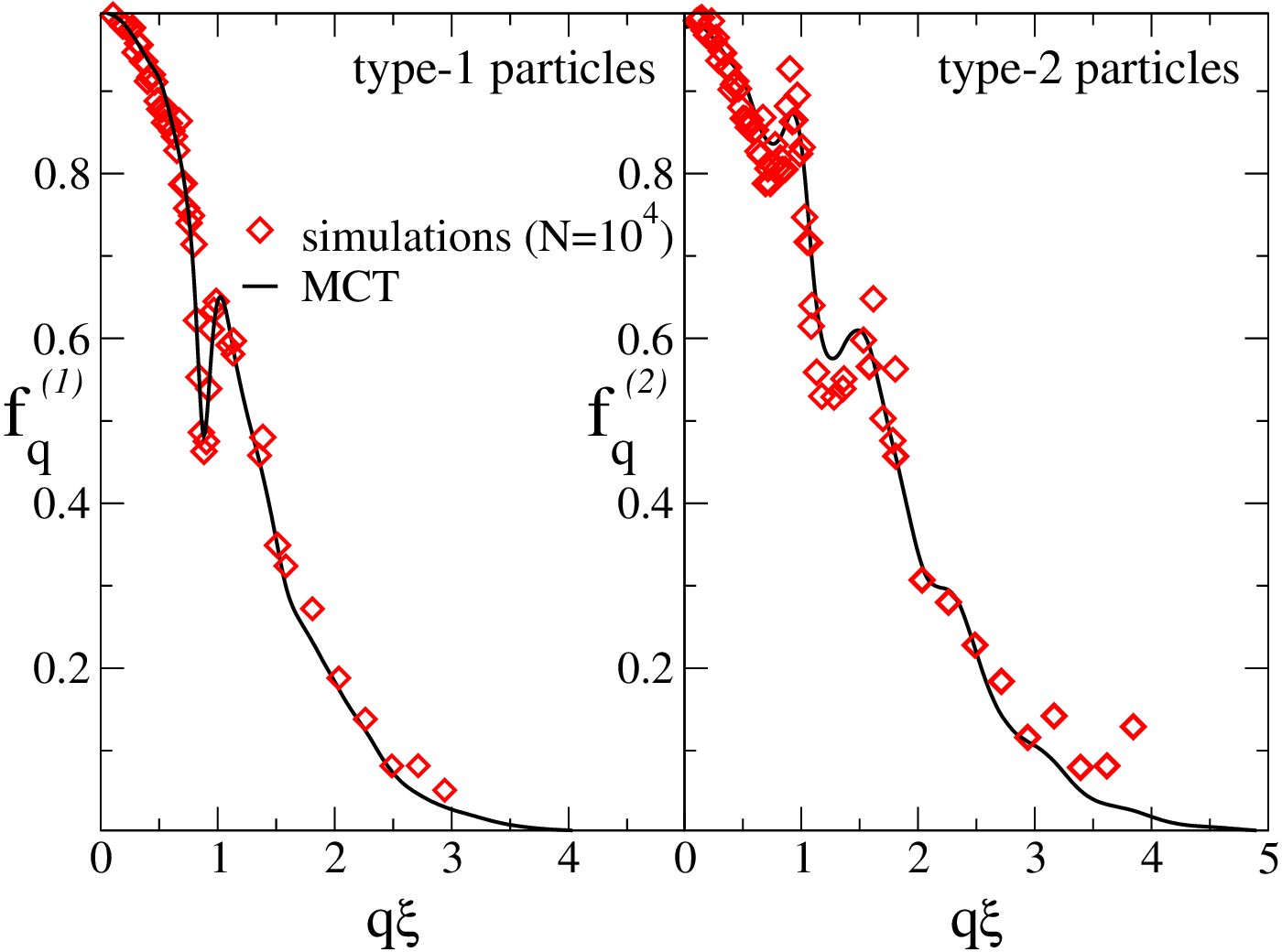}
\caption{
Debye-Waller factors from simulations extracted from stretched
 exponential fits (symbols) and from MCT calculations (lines) for
type-$1$ (left) and type-$2$ (right) particles respectively. The
simulation temperature is $T^*=2.0$, while for MCT calculations we
report results at the critical temperature $T^*_{MCT}=2.81$. }
\label{fig:fq}
\end{figure}

An interesting aspect of the MCT predictions\cite{Bos98a} is that the
two-step relaxation is maintained even as structural correlations
induced by Coulomb repulsion are diminished due to dilution
effects. In particular, the Debye-Waller factors,
$f_q^j=lim_{t\rightarrow\infty}\phi^{jj}(q,t)/S_{jj}(q)$ (where
$\phi^{ij}(q,t)$ is the collective intermediate scattering function)
for both species are expected to show weak oscillations as a function
of $q$ that become still weaker for lower density, until oscillations
are not observable\cite{Bos98a}.  While we could not access this
ultra-low density limit, the Debye-Waller factors extracted from our
simulations displayed in Fig.~\ref{fig:fq} show the precursor of this
behavior. Indeed, the simulation data show weaker oscillations in
$f_{q}^{j}$, as compared to that seen at $T_c^*$ for the standard
Kob-Andersen Lennard-Jones (KA) mixture \cite{KA,Nau97a}. Two other
features are noteworthy in the context of this comparison.  First, the
range of wave vectors over which caging is present is much smaller
than in standard high-density systems such as the KA mixture.  The
small $q$ values are reflective of caging at long length scales
induced by the soft long-ranged potential. Secondly, $f_q^j
\rightarrow 1$ for $q \rightarrow 0$.  This behavior is in contrast to
that seen for the majority species in the KA mixture\cite{Nau97a}, and
might result from the relatively large compressibility of the system
on these length scales\cite{ChenPRE}.
It should be noted that both features are
found in other soft materials such as colloidal gels\cite{Zac07a}.

We also calculated $f_{q}^{j}$ directly from MCT using the numerical
structure factors from Fig.~\ref{fig:sq}.
Critical MCT
Debye-Waller factors are also reported in Fig.~\ref{fig:fq} and they
are compared to the ones calculated from fitting the numerical
correlators for $T^* \geq T^*_c$.  The agreement is remarkable,
quantitatively matching all of the notable features exposed via direct
MD simulation.

Data presented in the previous figures show that a MCT description of the
dynamics can be  applied for $T^*>T^*_c$ and that the 
qualitative relaxation behavior is of the usual "type-B" variety\cite{goetze},
in agreement with  the predictions of Bosse and Wilke\cite{Bos98a,Wil99a}.
While we postpone a
detailed exposition for a
future publication, it should be noted that features such the
violation of the Stokes-Einstein relation, the growth of the
non-Gaussian parameter and multi-point dynamical susceptibilities also
occur in our simulated system.  These features become prominent close
to $T^*_{c}$ and behave in a manner similar to that found in standard
glass-forming systems.

In conclusion, we have clearly demonstrated the existence of a stable
classical Wigner glass state in a model of low-density charged
particles, which is independent of competing processes such as
crystallization or phase separation.  The low density soft glass
studied here exhibits significant differences from the high-density
hard-sphere glass. Most dramatically, we recall a much larger
localization length, which manifests itself both in the MSD plateau as
well as in the width of the Debye-Waller factors. Despite these
differences, several important aspects of the dynamics can be
successfully described by MCT.  On the other hand, deviations from
hard-sphere behaviour and from MCT predictions, in particular a small
value of the power-law exponent for the diffusion constant as well as
clear indications of activated processes, facilitated by the softness
of the potential, are also observed.
The possibility of reentrant relaxation emanating from an interplay of
hard core and long-ranged repulsion has not been explored in this
work, and will be the subject of future investigation.

EZ and FS acknowledge support from MIUR Prin and
MRTN-CT-2003-504712. DRR would like to thank NSF for support. We thank
K. Miyazaki for useful discussions.

\bibliographystyle{./apsrev}
\bibliography{./wigner}

\begin{thebibliography}{38}
\bibitem[{not()}]{notamixt}
\bibinfo{note}{We also investigated a different mixture with
  $\xi_{22}=1.5\xi_{11}=1.2\xi_{12}$, while $A_{ij}=\epsilon$ for each $i,j$.
  Results are fully qualitatively similar to the mixture with different
  amplitudes discussed in the text.}
\end{thebibliography}

\end{document}